\documentclass[aps,pre,twocolumn,showpacs,superscriptaddress]{revtex4}
\usepackage{graphicx}
\usepackage{amsmath,amssymb}
\usepackage{bm}

\newcommand{\ignore}[1]{}
\newcommand{\mbold}[1]{\mbox{\boldmath $ #1 $}}

\newcommand{\beq}{\begin{equation}}
\newcommand{\eeq}{\end{equation}}

\newcommand{\bxi}{\boldsymbol{\xi}}

\newcommand{\bH}{\boldsymbol{h}}

\newcommand{\cH}{\mathcal{H}}
\newcommand{\udd}{\text{d}}

\begin{document}

\title{The effect of the surface magnetic anisotropy of the neodymium atoms on the coercivity in the neodymium permanent magnet} 

\author{Masamichi Nishino}
\email{nishino.masamichi@nims.go.jp} 

\affiliation{Research Center for Advanced Measurement and Characterization, National Institute for Materials Science, Tsukuba, Ibaraki 305-0047, Japan}

\affiliation{Elements Strategy Initiative Center for Magnetic Materials, National Institute for Materials Science, 1-2-1 Sengen, Tsukuba, Ibaraki 305-0047, Japan}

\author{Ismail Enes Uysal}

\affiliation{Elements Strategy Initiative Center for Magnetic Materials, National Institute for Materials Science, 1-2-1 Sengen, Tsukuba, Ibaraki 305-0047, Japan}

\author{Seiji Miyashita} 

\affiliation{Department of Physics, Graduate School of Science, the University of Tokyo, 7-3-1 Hongo, Tokyo 113-0033, Japan}

\affiliation{The Physical Society of Japan, 2-31-22 Yushima, Tokyo 113-0033, Japan}

\affiliation{Institute for Solid State Physics, the University of Tokyo, 5-1-5 Kashiwanoha, Kashiwa 277-8581, Japan}

\affiliation{Elements Strategy Initiative Center for Magnetic Materials, National Institute for Materials Science, 1-2-1 Sengen, Tsukuba, Ibaraki 305-0047, Japan}

\date{\today}

\begin{abstract}
The Nd permanent magnet (Nd$_{2}$Fe$_{14}$B) is an indispensable material used in modern energy conversion devices. The realization of high coercivity at finite temperatures is a burning issue. 
One of the important ingredients for controlling the coercive force is the surface property of magnetic grains. 
It has been reported by first-principles studies that the Nd atoms in the first (001) surface layer facing the vacuum have in-plane anisotropy perpendicular to the $c$ axis, which may decrease the coercivity.  Focusing on the surface anisotropy effect on the coercivity, we examine the coercivity at zero and finite temperatures by using an atomistic model reflecting the lattice structure of the Nd magnet with a stochastic Landau-Lifshitz-Gilbert equation method. We study general three cases, in which the Nd atoms in surface layers have (1) no anisotropy, (2) in-plane anisotropy, and (3) reinforced anisotropy for two types of surfaces, (001) and (100) surfaces. 
We find that in contrast to the zero-temperature case, due to the thermal fluctuation effect, the modification of only the first surface layer has little effect on the coercivity at finite temperatures. However, the modification of a few layers results in significant effects. We discuss the details of the dependence of the coercivity on temperature, type of surface, and modified layer depth, and also the features of domain growth in magnetization reversal. 
\end{abstract}

\maketitle

\section{Introduction}
The realization of stronger coercivity at higher temperatures is a central issue in the development of permanent magnet materials for higher efficiency in energy conversion devices such as electric motors, generators, and electronic devices~\cite{Sugimoto}. The Nd magnet, Nd$_2$Fe$_{14}$B, is a very important target magnet because of its high coercive force~\cite{Sagawa,Herbst,Hirosawa1,Andreev,Kronmuller,Herbst2,Hirosawa2,O-Yamada,Mushnikov}. 

The coercivity depends on several factors, e.g., magnetic properties of grains (hard magnet phase) and grain boundaries (soft magnet phase)~\cite{Friedberg,Sakuma,Sakuma2,Mohakud,Wysocki,Uysal,Okamoto,Pramanik}. There a surface nucleation at the hard magnet is an essential process for magnetization reversal. Thus, the property of the surface of grains is a very important ingredient for the coercivity~\cite{Hirosawa_2017}. 
Indeed experimental enhancements of the coercivity by replacing surface Nd atoms by Dy atoms have been reported~\cite{Kim}. 

\begin{figure}[t]
  \begin{center}
     \includegraphics[width=6.5cm]{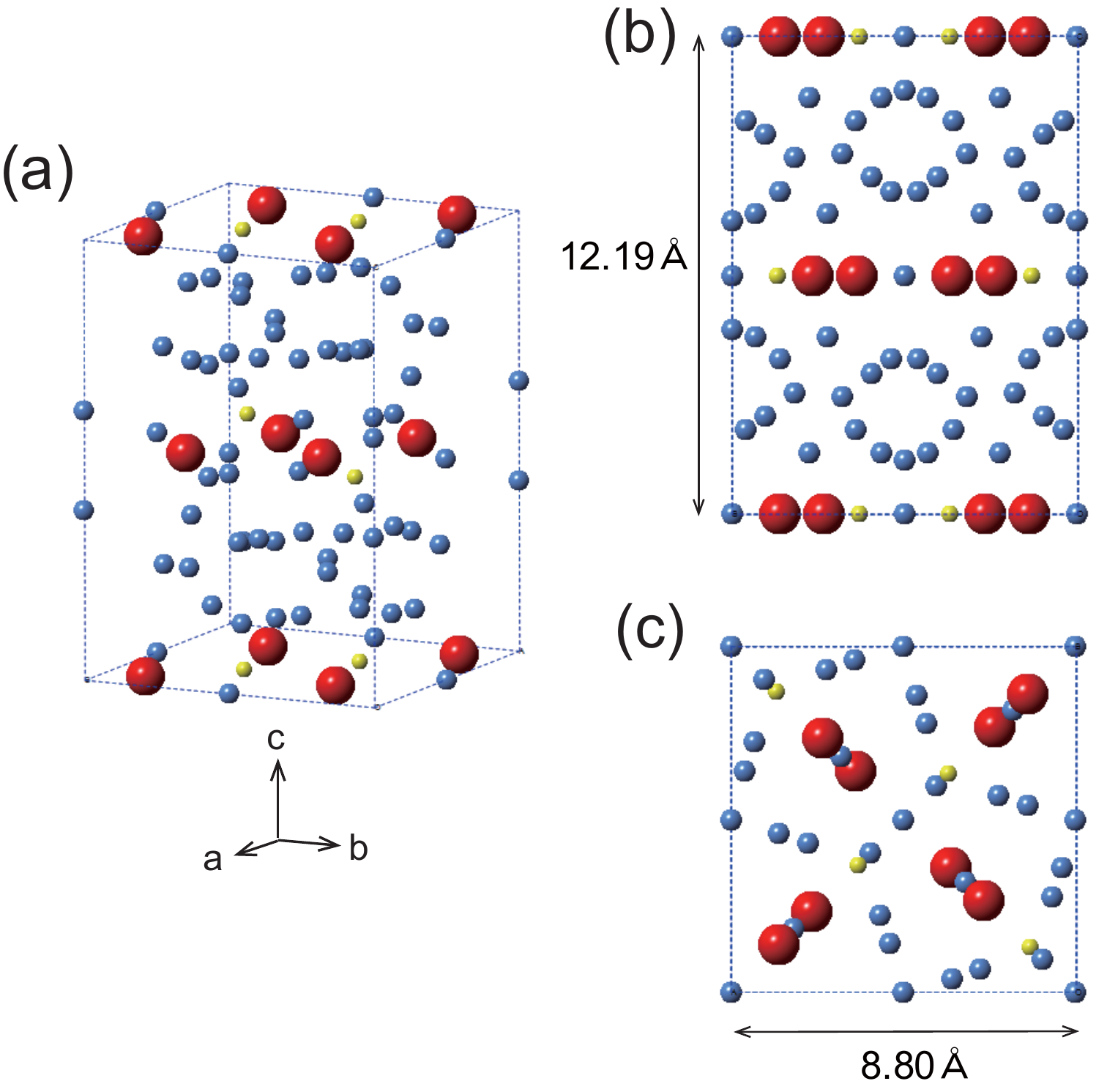}
  \end{center}
\caption{(a) Unit cell of Nd$_2$Fe$_{14}$B. Neodymium, Iron, and Boron atoms are denoted by red, blue, and yellow spheres, respectively. The lattice constants~\cite{Herbst} for the $a$, $b$, and $c$ axes are $d_{\rm a}=d_{\rm b}=8.80$ \AA, and $d_{\rm c}=12.19$ \AA, respectively. (b) Side view (from $a$ or $b$ axis). (c) Top view (from $c$ axis).}
\label{Fig_crystal_structure}
\end{figure}

Micromagnetics continuum modellings for permanent magnets have been applied in analyses of magnetic properties of the Nd magnet~\cite{book}. However, to study the microscopic details, recently developed atomistic models are indispensable~\cite{Toga,Nishino,Hinokihara,Miyashita,Toga2,Nishino2,Uysal,Toga3,Nishino3,Westmoreland,Westmoreland2,Gong1,Gong2,Gong3}. Unlike the continuum modellings, the lattice structure (Fig.~\ref{Fig_crystal_structure}) is introduced with microscopic magnetic parameters and the temperature effect can be treated properly in the atomistic modellings. By using the atomistic models, quantitative properties of the Nd magnet have been intensively investigated~\cite{Toga,Nishino,Hinokihara,Miyashita,Toga2,Nishino2,Uysal,Toga3,Nishino3,Westmoreland,Westmoreland2,Gong1,Gong2,Gong3}. 

The anisotropy energy for a rare-earth atom (ion) in the crystal electric field (CEF) is given by the following Hamiltonian: 
\begin{align}
{\cal H}_{\rm CEF} & = \sum_{l,m} B_{l}^m \hat{O}_{l}^m 
\label{CEF_Ham}
\end{align}
with $B_{l}^m =   \Theta_{l} A_{l}^m \langle r^l\rangle$. 
Here $B_{l}^m$ is the CEF coefficient and $\hat{O}_{l}^m$ is the Stevens operator, e.g., $\hat{O}_{2}^0=3J_z^2-J^2$, where $J=9/2$ for Nd atoms. 
 $\Theta_{l}$,  $A_{l}^m$, and $\langle r^l\rangle$  are the Stevens factor, the coefficient of the spherical harmonics of the crystalline electric field, and the average of $r^l$ over the radial wave function, respectively. 

The Nd magnet shows a spin-reorientation (SR) transition at $T_{\rm R}\sim150$ K, in which the magnetization is tilted from the $c$ axis at 0 K and becomes parallel to the $c$ axis above $T_{\rm R}$. 

\begin{figure}[th]
  \begin{center}
     \includegraphics[width=6.5cm]{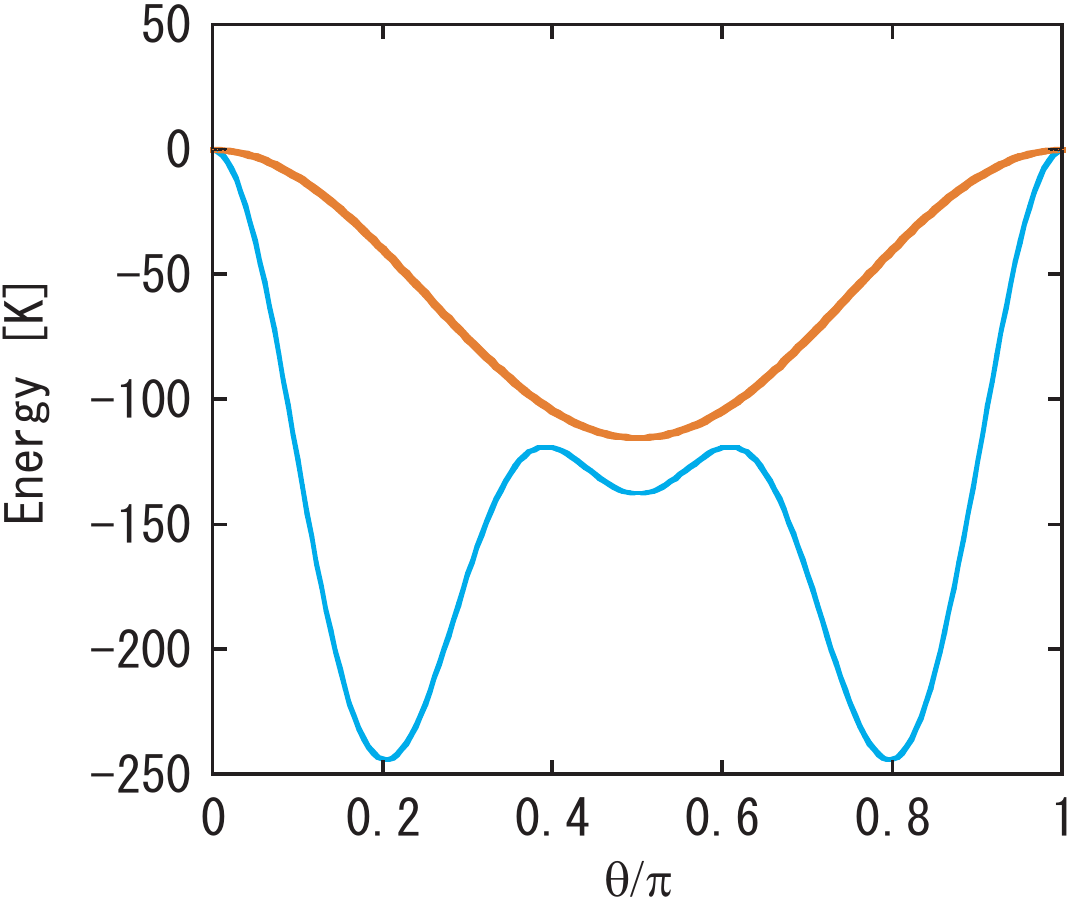}
  \end{center}
\caption{Anisotropy potential of the Nd atom (blue line) and in-plane anisotropy potential (red line) used in case (2) as functions of the angle $\theta$ from the $c$ axis. }
\label{Fig_potential}
\end{figure}

Substituting $J_z=J \cos \theta$ into Eq.~(\ref{CEF_Ham}), the anisotropy energy at 0 K for Nd atoms is expressed with diagonal terms in the following form: 
\beq
E_{\rm A}=K_1\sin^2\theta+K_2\sin^4\theta+K_4\sin^6\theta,
\label{Ea}
\eeq
where 
\beq
K_1=-3f_2B_2^0-40f_4B_4^0-168f_6B_6^0
\eeq
with constants $f_l(>0)$. 
The SR transition at $T_{\rm R}$ is well reproduced by the coefficients $A_{0}^2=295.0$ K $a_0^{-2}$,  $A_{0}^4=-12.3$ K $a_0^{-4}$, and $A_{0}^6=-1.84$ K $a_0^{-6}$ estimated by Yamada et al.~\cite{Yamada}, in which $a_0$ is Bohr radius. 
Although $A_{0}^2>0$ and $B_2^0<0$ ($\because$  $\Theta_{2}<0$), 
the first single ion-anisotropy $K_1<0$ due to the contribution of the other terms ($B_4^0$ and $B_6^0$ terms). This causes a tilted magnetization in the ground state (at zero temperature) as shown in Fig.~\ref{Fig_potential} (blue line), in which $\theta \simeq 0.2 \pi$ gives the minimum. 

\begin{figure}[th]
  \begin{center}
     \includegraphics[width=6.5cm]{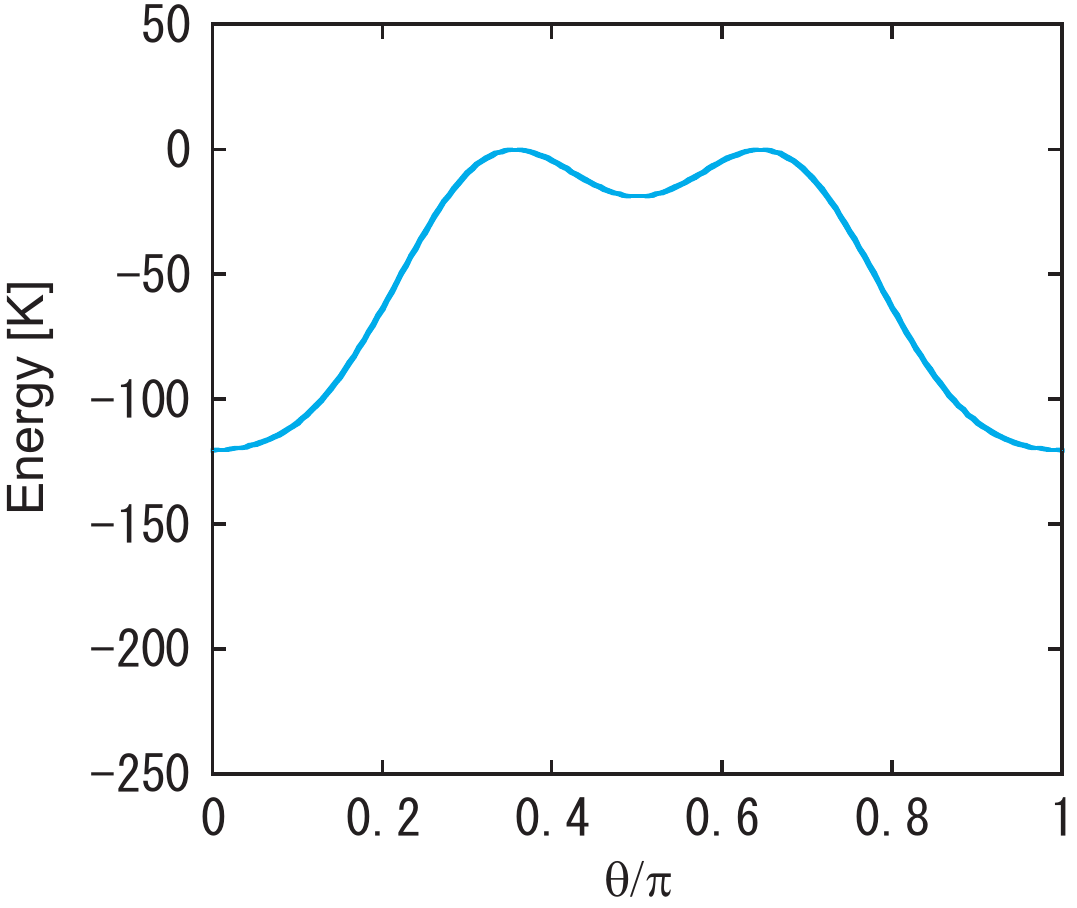}
  \end{center}
\caption{Effective anisotropy potential of the Nd atom at $T=0.46T_{\rm c}$ as a function of the angle $\theta$ from the $c$ axis.}
\label{Fig_potential_finite-T}
\end{figure}

At finite temperatures, the moment $J_z$ shrinks due to the thermal fluctuation as $J_z=CJ\cos \theta$ with $0<C<1$. 
The coefficient $C$ is estimated to be $C\simeq 0.76$ at $T=0.46 T_{\rm c}$ ($T_{\rm c}$ is the Curie temperature) close to room temperature in the bulk from the temperature dependence of the magnetization in our previous study~\cite{Nishino}. Thus substituting $J_z=0.76J\cos \theta$ into Eq.~(\ref{CEF_Ham}), the effective $K_1(T)$ in the form of Eq.~(\ref{Ea}) is found to be $K_1(T)>0$. 
Thus, for $T>T_{\rm R}$ the bottle-neck structure at 0 K is smeared out and a minimum at $\theta= 0$ is realized as shown in Fig.~\ref{Fig_potential_finite-T} (blue line).  

First-principles studies based on the density functional theory have reported that the Nd atoms in the first (001) surface layer have in-plane anisotropy perpendicular to the $c$ axis, different from out-of-plane anisotropy in the bulk, and this may cause a reduction of the coercivity~\cite{Moriya,Tanaka}. 
In the reports the estimated  $A_{0}^2$ shows $A_{0}^2<0$, i.e., $B_{0}^2>0$ for the Nd atoms at the (001) surface, which is different from that of the bulk, i.e., $A_{0}^2>0$, i.e., $B_{0}^2<0$, and the strength of $A_{0}^2$ for the surface is comparable to that in the bulk~\cite{Moriya,Tanaka}. 
Although the conclusion of these studies are derived within the first-term approximation, i.e., $K_1 \sim-3f_2B_2^0$, (namely $A_{0}^4=0$ and $A_{0}^6=0$), it indicates that the anisotropy at the surface is in-plane type. 

A large reduction of the coercive field was reported at zero temperature by a Landau-Lifshitz-Gilbert (LLG) simulation taking into account the in-plane anisotropy at the surface in a simplified lattice (cubic lattice) model~\cite{Mitsumata}. 
However, in this study the effect of tilted structure of the ground state was not considered. Furthermore, the temperature effect which may change the situation has not been treated. 
Thus, in the present paper, we study the surface magnetic anisotropy effect of Nd atoms in the Nd$_{2}$Fe$_{14}$B magnet on the coercive force by using a recently developed atomistic model constructed by the real crystal structure (Fig.~\ref{Fig_crystal_structure}). 
We investigate the temperature dependence of the surface effects for not only the (001) surface but also the (100) surface by employing the stochastic-LLG (SLLG) equation~\cite{Garcia,Nishino_SLLG}. 
We adopt $A_{0}^2$, $A_{0}^4$, $A_{0}^6$ given by Yamada et al. for the atomistic model in the bulk cells (see Sec.~\ref{model} and Fig.~\ref{systems} ). 
We focus on typical three cases for the surface anisotropy effect: (1) no anisotropy, (2) in-plane anisotropy, and (3) reinforced anisotropy for surface Nd atoms.

The rest of the paper is organized as follows. 
In Sec.~\ref{model}, the atomistic Hamiltonian, systems with different surface anisotropies, and the dynamical method are shown. 
In Sec.~\ref{threshold}, the coercive fields in the three cases for the (001) and (100) surfaces at zero temperature and finite temperatures are presented. 
Discussions about the surface anisotropy effects on the coercivity are also given. 
In Sec~\ref{Conf}, features of domain growth are discussed. 
Section~\ref{summary} is devoted to the summary. 

\section{Model and method}
\label{model} 

\subsection{Atomistic Hamiltonian for Nd$_2$Fe$_{14}$B}

We adopt the following atomistic Hamiltonian for the Nd magnet~\cite{Toga,Nishino,Hinokihara}: 
\begin{align}
{\cal H}=& - \sum_{i < j} 2 J_{ij} \mbold{s}_i \cdot \mbold{s}_j - \sum_i^{\rm Fe} D_i (s_i^z)^2   \\
& + \sum_{i}^{\rm Nd} \sum_{l,m} \Theta_{l,i} A_{l,i}^m \langle r^l\rangle_i \hat{O}_{l,i}^m -h \sum_i  S_i^z,  \nonumber \\ \nonumber
\label{model}
\end{align}
where $J_{ij}$ is the exchange interaction between the $i$th and $j$th sites, $D_i$ is the magnetic anisotropy constant for Fe atoms, the third term is the CEF (\ref{CEF_Ham}) for the magnetic anisotropy energy of Nd atoms, and $h$ is the external magnetic field applied to the $i$th site. The crystal structure of the unit cell is shown in Fig.~\ref{Fig_crystal_structure} (a). 
For Fe and B atoms, $\mbold{s}_i$ denotes the magnetic moment at the $i$th site, but for Nd atoms, it is  the moment of the valence (5d and 6s) electrons, and it is coupled antiparallel to the moment of the 4-f electrons $\mbold{\cal J}_i$. Thus the total moment for each Nd atom is $\mbold{S}_i=\mbold{s}_i + \mbold{\cal J}_i$. 
Here ${\cal J}_i =g_{\rm T}J \mu_{\rm B}$, in which $g_{\rm T}=8/11$ is Land\'e g-factor and $J=9/2$ is the magnitude of the total angular momentum. For the Fe and B atoms, we define $\mbold{S}_i=\mbold{s}_i$. 
In the third term the summation for $l$ runs $l=2,4,6$ and we consider only the diagonal operators ($m=0$) which give dominant contribution. 

We use the same parameter values for the atomistic model as in our previous studies~\cite{Toga,Nishino,Hinokihara}, in which the exchange interactions within the range of $r=3.52$ $\mathrm{\AA}$ for each atom were estimated by a first-principles calculation with the Korringa-Kohn-Rostoker (KKR) Green' s function method~\cite{Liechtenstein}, $D_i$ for Fe atoms (6 kinds) estimated in a first-principles study~\cite{Miura} were adopted, and for Nd atoms $A_{l}^m$ given by Yamada et al.~\cite{Yamada} were adopted with $\langle r^l\rangle$ estimated in Ref.~\cite{Freeman}. 
Employing these parameters, we showed the SR transition with transition temperature $T_{\rm R}= 150$ K, which is close to experimentally estimated ones~\cite{Hirosawa2,O-Yamada, Yamada,Mushnikov}, and the critical temperature $T_{\rm c} \sim 850$ K~\cite{Toga,Nishino}, which is a little overestimated to the experimental values $T_{\rm c} \sim$ 600 K~\cite{Hirosawa2,Andreev}. 

\begin{figure}[t]
  \begin{center}
     \includegraphics[width=9cm]{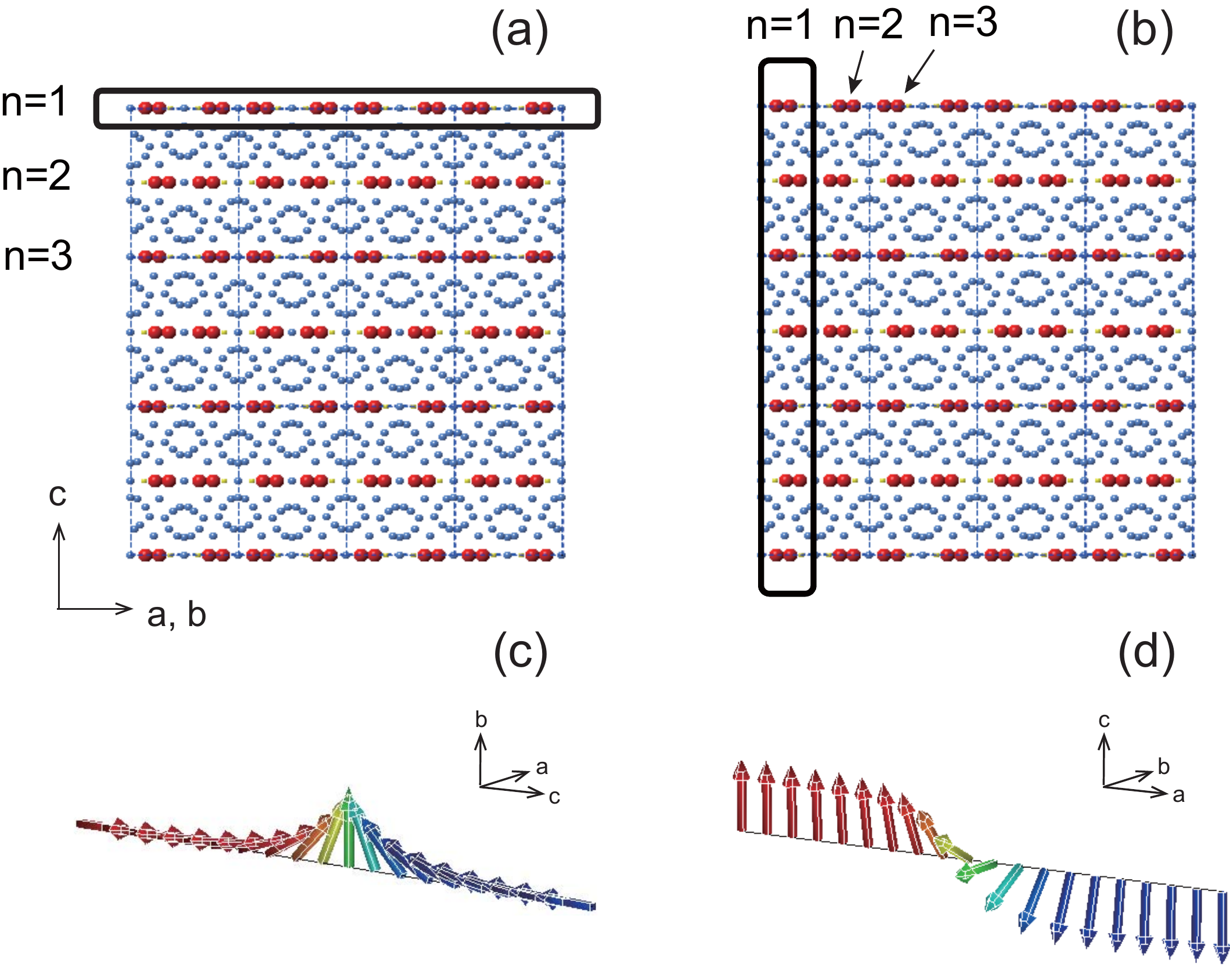}
  \end{center}
\caption{Definition of the Nd layer number for (a) the (001) surface and (b) the (100) surface. (c) Neel domain wall which moves along the $c$ axis. (d) Bloch domain wall which moves along the $a$ axis.}
\label{systems}
\end{figure}

\subsection{Systems}

Throughout this paper we study a system of 12 $\times$ 12 $\times$ 9 unit cells along the $a$, $b$, and $c$ axes, respectively (10.56 nm $\times$ 10.56 nm $\times$ 10.971 nm). We consider two kinds of surfaces, i.e., (001) and (100) surfaces as depicted in Fig.~\ref{systems}. 
The Nd layers at the (001) surface are numbered as in Fig.~\ref{systems} (a).
Nd atoms in the (100) surface at $n=$1, 2, $\cdots$ are defined as those located between $a=0$ and $a=d_{\rm a}/2$, between $a=d_{\rm a}/2$ and $a=d_{\rm a}$, $\cdots$, where $d_{\rm a}$ is the lattice constant of the $a$ axis (Fig.~\ref{systems} (b)). 
Open and periodic boundary conditions are used along the $c$ axis and the $a$ and $b$ axes, respectively, for the (001) surface, while open and periodic boundary conditions are used along the $a$ axis and the $b$ and $c$ axes, respectively, for the (100) surface. 
Thus, in each system two surface planes exist. 
A N\'eel type domain wall (DW) moves in the system with the (001) surface (Fig.~\ref{systems} (c)), while a Bloch type DW moves in that with the (100) surface (Fig.~\ref{systems} (d)). 

We investigate the coercive force in the following three cases of anisotropy parameters $({\tilde A}_2^0, {\tilde A}_4^0, {\tilde A}_6^0)$ for the surface Nd atoms. 

\begin{description}

\item [Case (1)] No anisotropy in Nd atoms: 
\beq
{\tilde A}_2^0={\tilde A}_4^0={\tilde A}_6^0=0, 
\eeq

\item [Case (2)] In-plane anisotropy in Nd atoms: 
\beq
{\tilde A}_2^0=-A_2^0<0, \quad {\rm and} \quad {\tilde A}_4^0={\tilde A}_6^0=0,
\eeq
where  $\tilde{A}_0^2$ is negative and the amplitude is the same order as that in the bulk ($A_0^2=295.0$K$a_0^{-2}$). 

\item [Case (3)] doubly reinforced anisotropy in Nd atoms: 
\beq
{\tilde A}_2^0=2A_2^0, \quad {\tilde A}_4^0=2A_4^0, \;\; {\rm and} \quad {\tilde A}_6^0=2A_6^0.
\eeq


\end{description}

\subsection{Dynamical method}

We study threshold fields under a reversed field by applying the SLLG equation\cite{Garcia,Nishino_SLLG}, 
\begin{equation}
\frac{\udd}{\udd t} \mbold{S}_i=-\frac{\gamma}{1+\alpha_i^2} \mbold{S}_i \times\bH_i^\text{\rm eff}-\frac{\alpha_i\gamma}{(1+\alpha_i^2)S_i}\mbold{S}_i\times\Big[\mbold{S}_i\times\bH_i^\text{\rm eff}\Big]. 
\label{EqLLG}
\end{equation}
Here the parameter $\gamma$ denotes the electron gyromagnetic ratio and $\alpha_i$ is the damping parameter.

The effective field $\bH_i^\text{\rm eff}$ on the $i$th spin consists of the exchange interaction field, the anisotropy field, the external field, and a noise field for thermal fluctuation, and it is given as 
\begin{equation}
\bH_i^\text{\rm eff} =
-\frac{\partial \cH}{\partial \mbold{S}_i} +\bxi_i(t). 
\end{equation}
Here the noise field $\bxi_i(t)=(\xi_i^x,\xi_i^y,\xi_i^z)$ is of white Gaussian and  satisfies the following relations:
\begin{equation}
\langle \xi_i^\mu(t)\rangle=0,\quad \langle \xi_i^\mu(t)\xi_j^\nu(s)\rangle=2 {\mathcal{D}}_i \delta_{ij}\delta_{\mu\nu}\delta(t-s). 
\end{equation}

The temperature $T$ is given by a function of the amplitude of the noise ${\mathcal{D}}_i$ according to 
the fluctuation dissipation relation:  
\begin{equation}
{\mathcal{D}}_i =\frac{ \alpha_i k_{\rm B}T } {\gamma S_i}. 
\end{equation}
With this relation, the system relaxes to a steady state (equilibrium) in the canonical distribution of the temperature $T$. 

Simulations were performed by solving Eq.~\eqref{EqLLG} numerically. $\alpha_i$ is set to 0.1. This choice does not affect the results in the present study because 
the value of $\alpha_i$ little affects the threshold field~\cite{Nishino3}. 
We employ a kind of middle-point method~\cite{Nishino_SLLG} equivalent to the Heun method for the numerical integration. For the time step of this equation, $\Delta t = 0.1$ fs is used. 


\begin{figure*}[t!]
  \begin{center}
     \includegraphics[width=12cm]{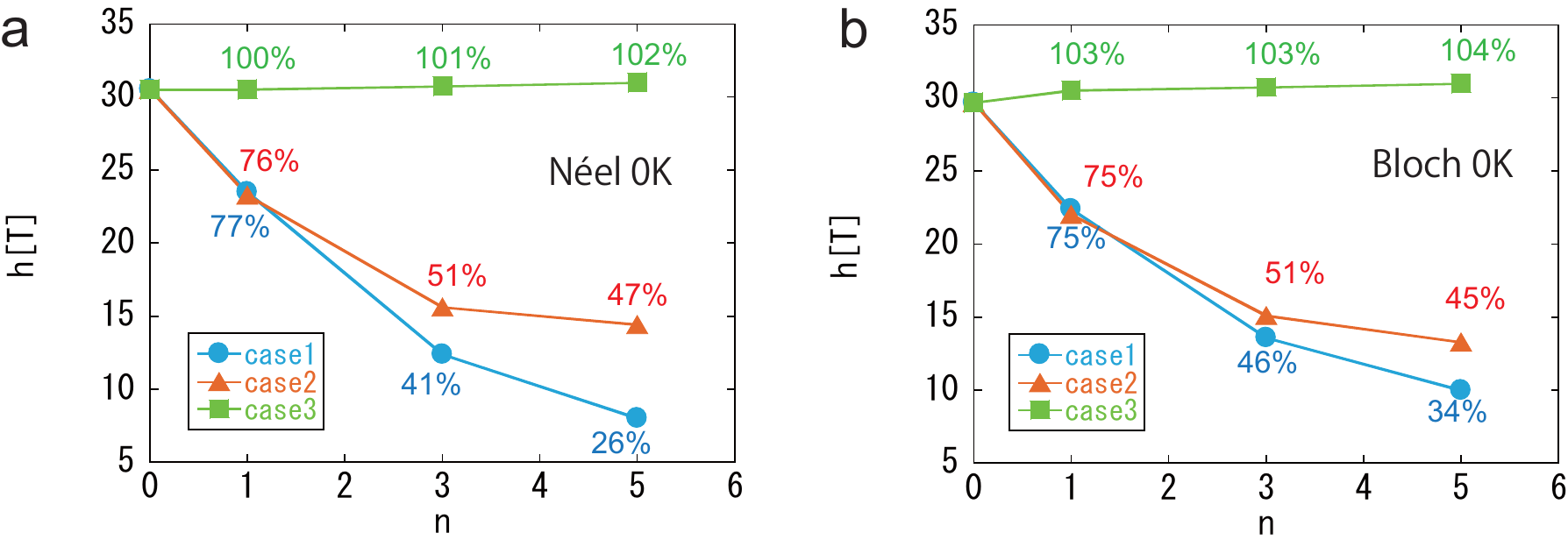}
  \end{center}
\caption{$n$ dependencies of the threshold fields at 0 K in cases 1--3 for (a) the  (001) surface and (b) the (100) surface. The values of percentage are the ratios of the threshold fields for $n$ to those for $n=0$.}
\label{n_vs_h_zero-temp}
\end{figure*}

Staring from a down-spin state for all spins, 
we calculated the time evolution of the magnetization 
\beq
M_z ={\sum_iS_{i,z}}
\eeq
under a given value of magnetic field $h$ in the $c$ direction. 
We define the time of reversal as the time when the magnetization changes its sign.  
At zero temperature, there is no thermal fluctuation, and thus when the magnetic field reaches the value to vanish the potential barrier, this value gives the threshold field for magnetization reversal, i.e., coercive field. This process is deterministic. 
We regarded the final state as the state when all the motion stops.  
In the simulation, the final state was obtained within $N_t =1\times 10^6$ time steps ($t=0.1$ ns). 
 
In contrast to zero-temperature reversal, at finite temperatures the magnetization reversal exhibits a stochastic process to jump over the energy barrier by thermal fluctuation. 
We defined the threshold magnetic field at finite temperatures as follows.
We performed twelve simulations with different random number sequences. 
At each value of the magnetic field, we counted the number $N$ of cases in which magnetization reversal took place within a simulation time $t_{\rm max}$. 
If $N=0$, the field is smaller than the threshold field, while if $N=12$, the field is larger than the threshold field. 
We defined the threshold field as the middle point of the interval between the field of $N=0$ and that of $N=12$. 
The error bar for the threshold field was defined as the transient region of the field. 
The threshold field depends on the simulation time. Here, we set the simulation time to be $t_{\rm max}=0.5$ ns ($5\times 10^6$ time steps). 
The measurement time for the coercivity should be order of 1s in experiments and this time scale is not practical for real-time simulations. 
However, the dependence of the reversal time on the external field around the threshold field is very sharp, i.e., the reversal time exponentially increases.
In our previous paper~\cite{Nishino3}, it was estimated that the threshold field for 1s is about 25 \% less than the value of the simulation with 0.5ns. 
Thus we expect that the estimated threshold fields can give approximated values for the coercive field. 


\begin{figure*}[t!]
  \begin{center}
     \includegraphics[width=12cm]{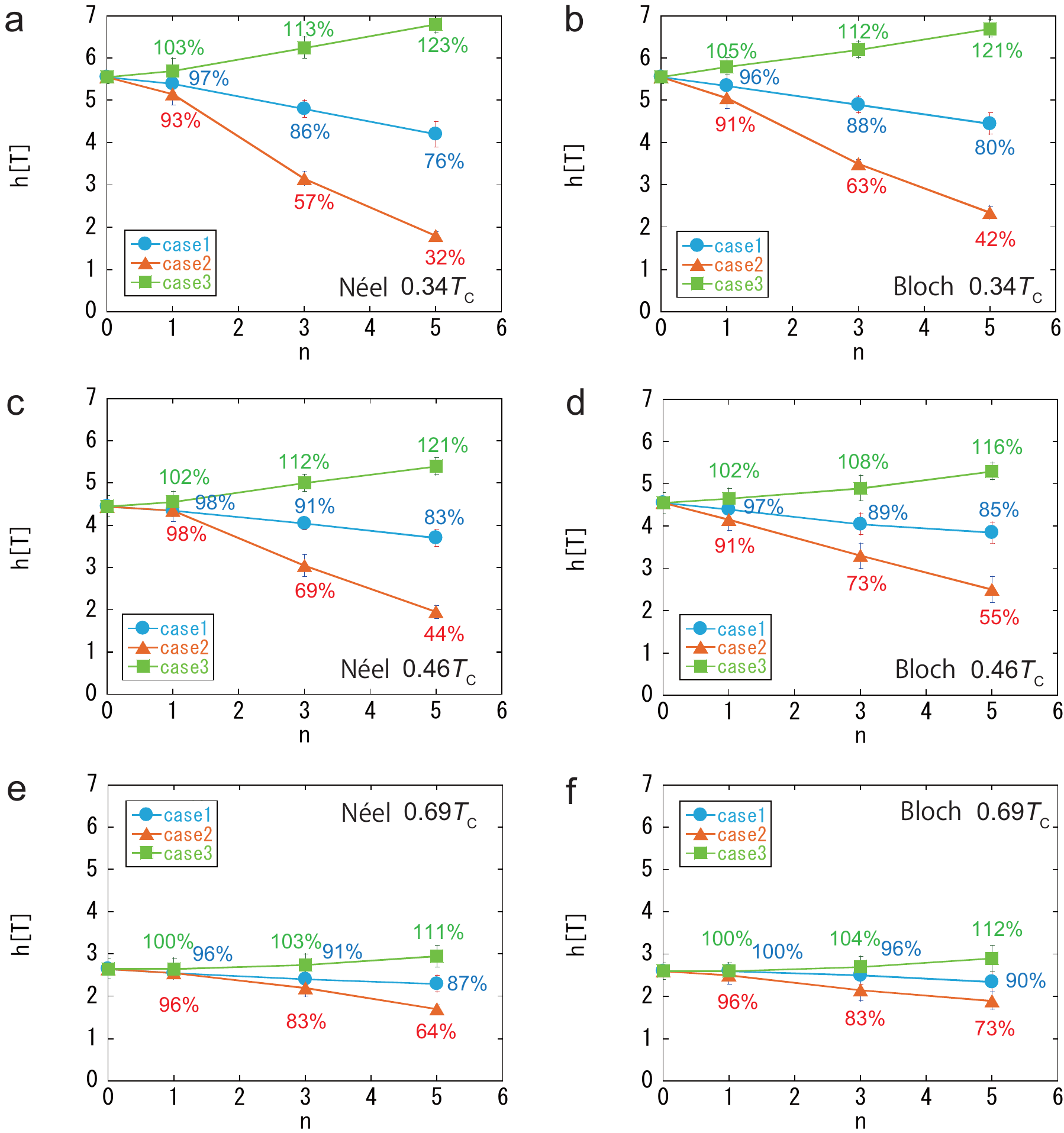}
  \end{center}
\caption{$n$ dependencies of the threshold fields in cases (1)--(3) for the  (001) surface at (a) $T=0.34T_{\rm c}$, (c) $T=0.46T_{\rm c}$, (e) $T=0.69T_{\rm c}$ and the (100) surface at (b) $T=0.34T_{\rm c}$, (d) $T=0.46T_{\rm c}$, (f) $T=0.69T_{\rm c}$. 
The values of percentage are the ratios of the threshold fields for $n$ to those for $n=0$.}
\label{n_vs_h_finite-temp}
\end{figure*}

\section{Threshold fields for magnetization reversal}
\label{threshold}

\subsection{Zero-temperature properties}

First, we study the threshold field at $T=0$ K.
In Figs.~\ref{n_vs_h_zero-temp} (a) and (b), we depict how the threshold field depends on the number of modified layers $n$ in the three cases (1), (2), and (3) for both cases of the (001) and (100) surfaces, respectively. 
Here $n=0$ is defined as the case without modification of the anisotropy of surface Nd atoms, i.e., the anisotropy of the surface Nd atoms is the same as that in the bulk. The value (percentage) given for each symbol in the figures denotes the ratio of the threshold field to that for $n=0$. 

We find that for both surfaces, the modification of the single surface ($n=1$) causes a large amount of reduction of the threshold field, i.e., around 25 \% in cases (1) and (2), and the field decreases further for larger $n$. 
This indicates that at $T=0$ K only one-layer modification of Nd atoms affects the coercive field largely. 

It should be noted that the reduction of the threshold fields in case (1) (blue circles) is larger than that in case (2) (red triangles). This means that the in-plane anisotropy helps the coercivity in comparison with the case of no anisotropy.
This tendency is different from that at finite temperatures studied in the next subsection. We consider the cause of this difference in the next subsection.

The threshold field in case (1) for the (001) surface is smaller than that for the (100) surface ($n \ge 3$). This may be due to weaker exchange interactions along the $c$ axis than those along the $a$ axis~\cite{Nishino,Toga2}. 
A similar tendency was also seen in the nucleation field from a soft magnet phase in our previous study, in which the nucleation field for the Neel DW is smaller than that for the Bloch DW~\cite{Uysal}. 

In case (3), the threshold field changes with $n$, e.g., 2\% for $n=5$ at the (001) surface and 4 \% for $n=5$ at the (100) surface. 
This small increment is different from the dependence at finite temperatures. 
It is worth noting that for $n=0$ the values of the threshold fields are very high in both cases of the (001) and (100) surfaces. 
It is considered that for $n=0$ the magnetization reversal occurs in the Stoner-Wohlfarth mechanism in which the whole magnetic moments rotate simultaneously. Therefore, even if the surface anisotropy is reinforced, the reversal occurs in the bulk and the reinforcement effect is small. 

\subsection{finite-temperature properties}

In Fig.~\ref{n_vs_h_finite-temp}, we depict $n$ dependencies of the threshold fields in cases (1)--(3) for the (001) surface at (a) $T=0.34T_{\rm c}$, (c) $T=0.46T_{\rm c}$, (e) $T=0.69T_{\rm c}$,
 and those for the (100) surface (b) at $T=0.34T_{\rm c}$, (d) $T=0.46T_{\rm c}$, (f) $T=0.69T_{\rm c}$. 

In contrast to the threshold field at 0 K, one-layer modification for Nd atoms ($n=1$) affects little the threshold fields at these finite temperatures in all the three cases ((1)--(3)) for both surfaces. 
For example, at $T=0.46T_{\rm c}$ which is close to room temperature, the threshold fields are larger than 90\% of those for $n=0$ for the (001) and (100) surfaces in cases (1) and (2). 
This result indicates that the thermal fluctuation smears out the effect of the surface in the case of $n=1$. 

However, the modification gives relevant effects on the threshold field as $n$ increases. As the temperature is raised, the threshold field for $n=0$ decreases and the effect of the modification becomes weaker because the thermal fluctuation becomes stronger and further smears the surface effect. 

In contrast to the observation at $T=0$ K, at finite temperatures we find that the threshold fields in case (2) are smaller than those in case (1), which indicates that the in-plane anisotropy accelerates magnetization reversal with the thermal fluctuation. 
We also confirmed this property at $T=0$ K in a simplified system of an open chain, in which Nd atoms are connected with an exchange interaction, and only a few Nd atoms at one end have the anisotropy potential of case (1) or case (2), and the others have the anisotropy in the bulk. 
We performed a calculation to find the minimum energy state including the metastable state at a given field, and estimated the threshold field for magnetization reversal. The result shows that in a wide region of the strength of the exchange interaction, the coercive force in case (2) is larger than that in case (1). This property is interpreted as follows. 

The minimum point of the anisotropy potential energy of the Nd atoms in the bulk is located at $\theta \simeq 0.2 \pi$ from the $c$ axis as shown in Fig.~\ref{Fig_potential} (blue line). On the other hand, for the Nd atoms in the modified layers in case (2), the anisotropy potential energy has a minimum at $\theta=\pi/2$ (Fig.~\ref{Fig_potential} (red line)). 
Around the border between the modified and unmodified layers, the Nd magnetic moments in the modified layers point at some $\theta$ smaller than $\theta=\pi/2$ due to the effect from the bulk, while those in the unmodified layers point at some $\theta$ larger than $\theta \simeq 0.2 \pi$ due to the effect from the modified Nd atoms. 
This compromise reduces the loss of the exchange energies and leads to 
a more stabilized configuration in case (2) than that in case (1) as shown in the threshold fields of Fig.~\ref{n_vs_h_zero-temp}. 

Contrary to zero temperature, in Fig.~\ref{n_vs_h_finite-temp} we find that case (1) is more stable than case (2). At finite temperatures the effect of the thermal fluctuation plays an important role in magnetization reversal. 
The anisotropy of the Nd atoms in the bulk is parallel to the $c$ axis at finite temperatures, and thus the in-plane anisotropy causes a large thermal fluctuation of the magnetizations due to the frustrated situation. This fluctuation causes the Fe moments in the modified layers and also in the vicinity of the modified layers to much fluctuate. This results in the acceleration of the magnetization reversal.  

In case (3), in contrast to the zero-temperature case, the threshold fields at finite temperatures increase for larger $n$, which is naturally understood due to the reinforcement effect of the anisotropy parallel to the $c$ axis. 

In all three cases, the effect of the modification of the surface Nd atoms is larger for the (001) surface than for the (100) surface. 
This is probably because exchange interactions along the $c$ axis is weaker than the $a$ ($b$) axis, and thus the modification effect of the anisotropy energy for the (001) surface appears relatively stronger than that for the (100) surface.

\begin{figure}[t!]
  \begin{center}
     \includegraphics[width=9cm]{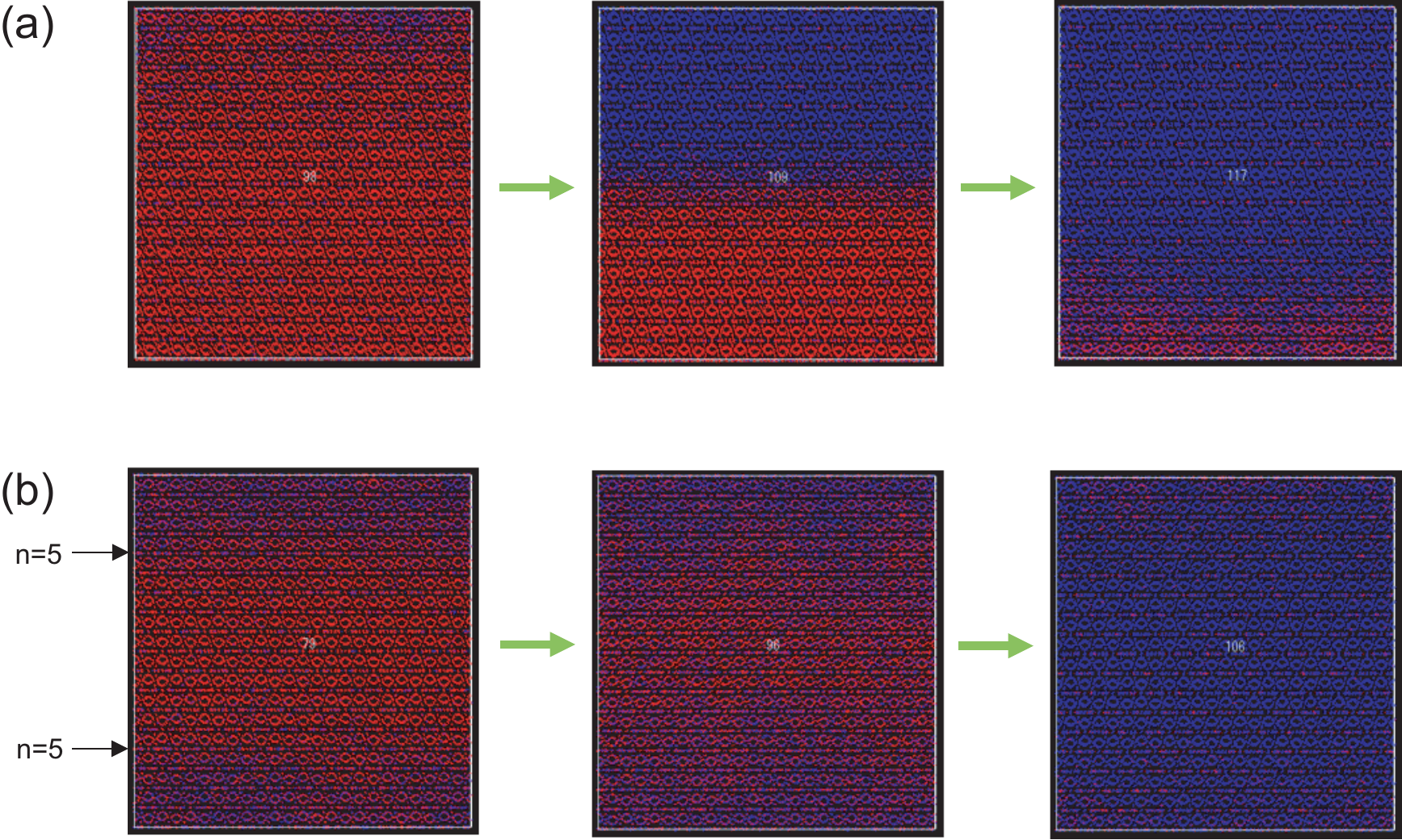}
  \end{center}
\caption{Snapshots of the magnetic moments for the (001) surface at  $T=0.34T_{\rm C}$ for (a) $n=0$ at $h=5.5$ T and (b) $n=5$ at $h=1.8$ T. The left panel of each figure is a snapshot for starting of magnetization reversal, the middle panel is that in the middle of the reversal, and the right panel is that of the last stage of the reversal. Red and blue denotes down-spin ($S_i^z<0$) and up-spin states ($S_i^z>0$), respectively.}
\label{Snapshot_001}
\end{figure}

\begin{figure}[t!]
  \begin{center}
     \includegraphics[width=9cm]{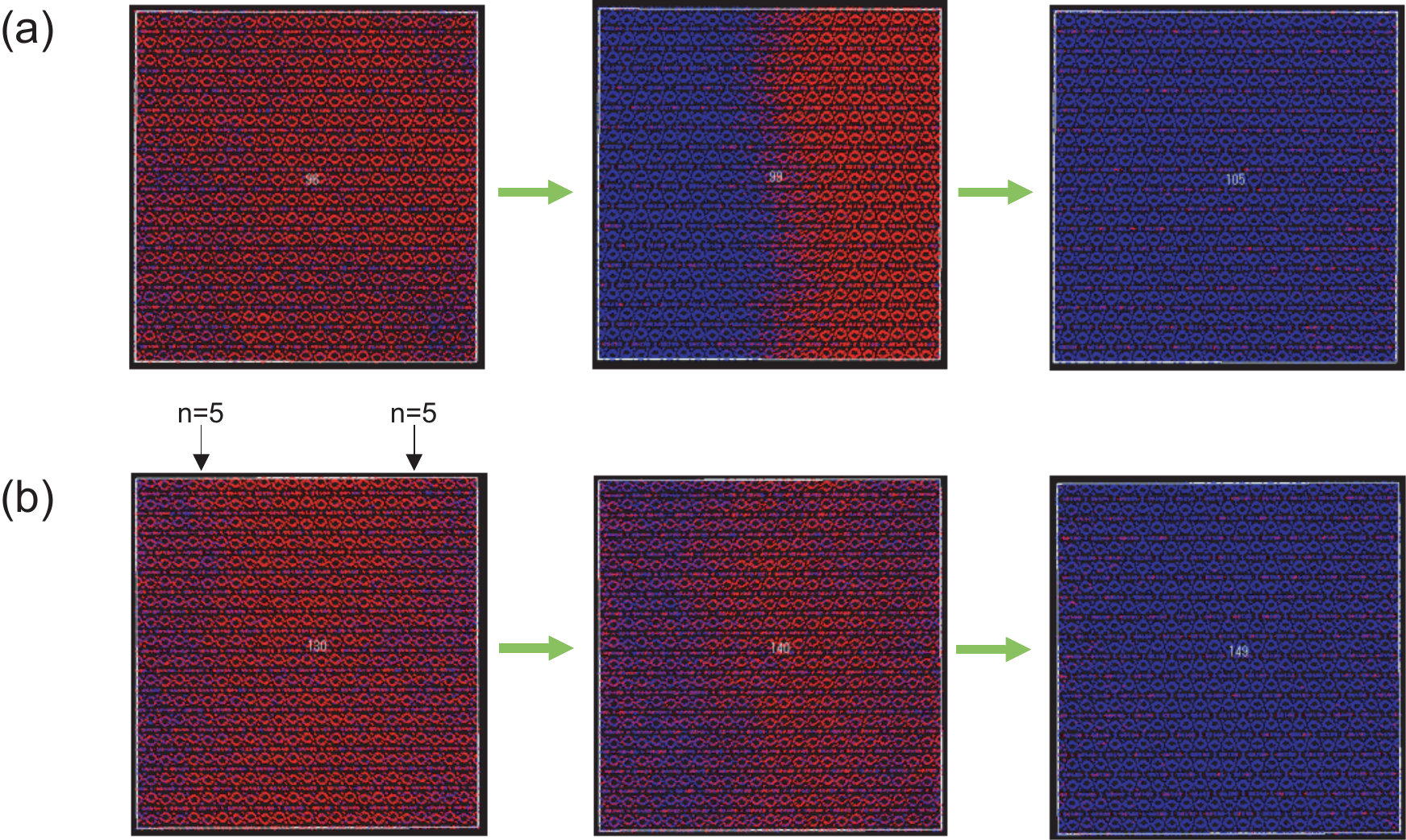}
  \end{center}
\caption{Snapshots of the magnetic moments for the (100) surface at  $T=0.34T_{\rm C}$ for (a) $n=0$ at $h=5.5$ T and (b) $n=5$ at $h=2.3$ T. The left panel of each figure is a snapshot for starting of magnetization reversal, the middle panel is that in the middle of the reversal, and the right panel is that of the last stage of the reversal.}
\label{Snapshot_100}
\end{figure}

\section{Configurations in the reversal process} 
\label{Conf}

To catch the feature of the domain growth in the magnetization reversal, we investigate snapshots of magnetic configurations. 
We give in Fig.~\ref{Snapshot_001} snapshots of the configuration of the magnetic moments in the system with the (001) surface at  $T=0.34T_{\rm c}$ for (a) $n=0$ at its threshold field $h=5.5$ T, and (b) $n=5$ in the case (2) at its threshold field 
$h=1.8$ T (see also animation\_001\_a and animation\_001\_b in supplementary materials for detailed features~\cite{suppl}). 
The left panel is a snapshot for starting of magnetization reversal, the middle panel is that of domain wall propagation, and the right panel is that of the last stage of the reversal. 
In the case $n=0$, before magnetization reversal, a small up-spin domain appears at the right side around the surface in the left panel of Fig.~\ref{Snapshot_001} (a). Similarly small up-spin domains appear and disappears around the surface (see also animation\_001\_a in supplementary materials). Then, a nucleation occurs near the surface and a domain grows along the $a$ and $b$ axes quickly. The domain moves parallel to the $c$ axis. 
On the other hand, for $n=5$ before starting of the magnetization reversal, not only the Nd magnetic moments but also a part of the Fe magnetic moments located between the first and fifth layers fluctuate much larger than those in the bulk because the Fe moments are affected by the fluctuation of the surface Nd moments through the exchange interactions between the Nd and Fe atoms. Thus the domain formation takes place more easily and it grows parallel to the $c$ axis. It is noting that the border between up and down-spin parts is more unclear in Fig.~\ref{Snapshot_001} (b)  than in Fig.~\ref{Snapshot_001} (a) although the external field is weaker. This is probably because affected by those fluctuating moments the Fe moments in the bulk layers also fluctuate through strong exchange interactions.  

We also show in Fig.~\ref{Snapshot_100} snapshots of the magnetic moments for the (100) surface at  $T=0.34T_{\rm c}$ for (a) $n=0$ at its threshold field $h=5.5$ T and (b) $n=5$ in the case (2) at its threshold field $h=2.3$ T (see also animation\_100\_a and  animation\_100\_b in supplementary materials for detailed features). 
Before magnetization reversal for $n=0$, up-spin domains appear and disappear around the surface in similar manner to the observation of the (001) surface but the domains are larger and fluctuation range is larger (see also animation\_001\_b in supplementary materials). This is partially because the width of the Bloch DW is larger than that of the N\'eel DW~\cite{Nishino}. After nucleation, the domain moves parallel to the $a$ axis. 
For $n=5$, the Nd moments and a part of the Fe moments near the surface located up to around the fifth layer fluctuate larger than those in the bulk, but the range of the fluctuation is not so clear as that for the (001) surface. This is probably because layered structure exists along the $c$ axis and also the DW width is larger than that of the case for the (001) surface.  
Similarly in Fig.~\ref{Snapshot_001}, the border between up and down-spin parts is more unclear in Fig.~\ref{Snapshot_100} (b) than in Fig.~\ref{Snapshot_100} (a).

\section{Summary}
\label{summary}

We investigated how the modification of magnetic anisotropy of the Nd atoms located near the surface affects the coercivity at zero and finite temperatures by applying the stochastic LLG equation to the atomistic model.  
Reflecting the lattice structure, we studied the two cases of the surfaces, i.e., 
the (001) surface and the (100) surface. 
We examined the three cases of the modification, i.e., the Nd atoms in surface layers have (1) no anisotropy, (2) in-plane anisotropy, and (3) doubly reinforced anisotropy. We showed the temperature dependence and the modified-layer-depth dependence of the coercivity. 

At zero temperature $T=0$ K, one-layer modifications of the anisotropy of the Nd atoms in cases (1) and (2) largely reduce the coercivity. The reduction in case (1) is larger than in case (2), in which the in-plane anisotropy helps to maintain the coercivity compared to no anisotropy. The reinforcement little affects the coercivity at $T=0$ K. 
In contrast to $T=0$ K, at finite temperatures above the SR transition point, the one-layer modifications hardly affect the coercive field in all three cases. This revises the result of $T=0$ K. 
Namely, the thermal fluctuation effect plays an important role in the magnetization reversal at finite temperatures. 

As the number of the modified layers $n$ increases, the effect of modification becomes large. Unlike the zero-temperature case, at finite temperatures the reduction of the coercivity in case (1) is smaller than in case (2). Namely, the in-plane anisotropy more suppresses the reversal at $T=0$ K than the no anisotropy case, while it accelerates the reversal at finite temperatures. 
The difference is attributed to the change of the structure of the anisotropy potential energy between zero (the ground state) and finite temperatures above the SR transition point. 

The coercivity increases significantly in case (3) at finite temperatures, which indicates that the surface treatment with strong anisotropic atoms would help the coercivity to increase. Unlike this situation, the coercivity little changes at $T=0$ K in case (3), which is due to the uniform rotation. 

In all the three cases at finite temperatures, the effect of the modification of the surface Nd atoms is larger for the (001) surface than for the (100) surface for larger $n$, which is probably due to weaker effective exchange interactions along the $c$ axis than the a (b) axis. 

We also investigated features of the domain formation and propagation. 
As to the (001) surface without modification at finite temperatures, a nucleation occurs suddenly near the surface and a domain grows along the $a$ and $b$ axes quickly and a domain wall moves along the $c$ axis. On the other hand, for $n$ layer modification in case (2), the magnetic moments of not only the Nd atoms but also the Fe atoms located up to $n$ surface layer fluctuate largely before the magnetization reversal, and it makes a nucleation occur more easily. 
In the case of the (100) surface, a similar feature is observed but magnetic domains fluctuate in wider range regardless of the $n$ layer modification, which may be partially due to a larger domain-wall width along the $a$ ($b$) axis. 

Experimentally the coercive field of the Nd magnet has been enhanced by adding Dy atoms. It has been recently reported that an annealing process after the diffusion made Dy-rich second shells whose widths are a few nanometers, and those shells enhance the coercive force by 30 to 40 \%~\cite{Kim}. 
Our present study can give useful microscopic information to such related experiments. 
The effect of the replacement of surface Nd atoms by Dy atoms on the coercivity is a challenging future subject, which requires atomistic information of the magnetic parameters of the replacement by Dy atoms. It will be studied in the future. 

\section*{Acknowledgments}
The authors would like to thank Dr. Hirosawa for instructive discussion from experimental view points, and Dr. Toga and Dr. Hinokihara for useful discussion for theoretical aspects. The present work was supported by the Elements Strategy Initiative Center for Magnetic Materials (ESICMM) (Grant No. 12016013) funded by the Ministry of Education, Culture, Sports, Science and Technology (MEXT) of Japan, and was partially supported by Grants-in-Aid for Scientific Research C (No. 18K03444 and No. 20K03809) from MEXT. The numerical calculations were performed on the Numerical Materials Simulator at the National Institute for Materials Science.

\end{document}